# Many-body effect in optical properties of monolayer molybdenum diselenide.

Ke Xiao[1], Tengfei Yan[1,*], Qiye Liu[1], Siyuan Yang[1], Chiming Kan[1], Ruihuan Duan[2], Zheng Liu[2], Xiaodong Cui[1,*]

[1] Department of Physics, University of Hong Kong, Hong Kong SAR
[2] School of Materials Science and Engineering, Nanyang Technological University, Singapore 639798, Singapore

* e-mail: yantf@hku.hk and xdcui@hku.hk

Abstract: Excitons in monolayer transition metal dichalcogenide (TMD) provide a paradigm of composite Boson in 2D system. This letter reports a photoluminescence and reflectance study of excitons in monolayer molybdenum diselenide ($MoSe_2$) with electrostatic gating. We observe the repulsive and attractive Fermi polaron modes of the band edge exciton, its excited state and the spin-off excitons. Our data validate the polaronic behavior of excitonic states in the system quantitatively where the simple three-particle trion model is insufficient to explain.

Two-dimensional (2D) transition metal dichalcogenide (TMD) is being recognized as a new emulation platform for many body effects in 2D system. The intrinsic 2D nature dramatically enhances the Coulomb interactions as is evidenced by the giant exciton binding energy in monolayer TMDs of a few hundred meVs which is at least one order of magnitude higher than that in conventional quasi-2D systems based on III-V semiconductor heterostructures.[1-4] Apart from a direct bandgap in visible range and strong optical oscillator strength,[5,6] nonzero Berry curvature,[7,8] valley dependent optical selections and spin-valley locking provide monolayer TMDs rich features in physics.[9-13] One paradigm is the excitonic effect in monolayer TMDs. Excitons in monolayer TMDs are of Wannier-Mott type, being delocalized with a Bohr radius of ~1nm, despite with a Frenkel type like giant binding energy owing to enhanced Coulomb interaction.[14,15]

Excitons as a quasiparticle of Boson submerged in electric tuned charge carriers in monolayer TMDs provide a solid state paradigm of an 'impurity problem' in 2D.[16] The concept of Fermi polaron is applied to describe such many-body interactions on excitons in monolayer TMDs where diluted excitons acting as atomic impurities immersed in Fermi sea of charge carreirs.[17] Besides the vast studies in solids, polarons are investigated in cold atom systems in recent years, in which minority atoms act as impurities under majority atom bath.[18-22] In contrast to cold atom system where the interaction strength is tuned with an external magnetic field near Feshbach resonance and the quasi-particle, polaron, is usually investigated with radio-frequency spectroscopy,[18,20,21] in 2D TMDs, the interaction is orders of magnitude bigger and could be manipulated by an external electric field.[23] The prominent optical response of 2D TMDs is dominated by the interband excitons owing to their giant binding energy and optical oscillator strength. Under the presence of charge carriers, excitons are easily bound to carriers (electrons or holes) with Coulomb interaction to form a new quasiparticle state, charged bound excitons, or so called trions. This was described with a simple trion model of a bound complex of one electron (hole) and one exction.[24] This model works well at low electron/hole density, whereas not so well at elevated carrier density. The model of Fermi polaron (eq.1) is thus proposed and shows a good consistence with the experiments.[17, 23, 25-27]

$$\hat{H} = \sum_k \varepsilon_{e,k} \hat{c}_k^\dagger \hat{c}_k + \sum_K \varepsilon_{X,K} \hat{X}_K^\dagger \hat{X}_K + \sum_{QKq} U_{QKq} \hat{c}_{Q-q}^\dagger \hat{c}_Q \hat{X}_{K+q}^\dagger \hat{X}_K \quad (1)$$

The first two terms are the electron and exciton kinetic energy which can be written as $\varepsilon_{e,k} = \frac{k^2}{2m_e}$ and $\varepsilon_{X,K} = \varepsilon_{X,0} + \frac{K^2}{2m_X}$. The third term describes the interaction between excitons with electrons in Fermi sea. The collective scattering involves Fermi sea fluctuations where transient electron-hole pairs, scattered electron/hole and the corresponding vacancy state in conduction/valence band, are generated, analogous to phonon polaron where electrons are dressed by phonons (lattice vibration) in crystals. These collective interactions lead to two distinct quasiparticle modes, attractive mode at the lower energy side and repulsive mode at the higher which are recognized as trion and free exciton states near charge neutral region, respectively. (fig.1 a and b).

Here we report the photoluminescence (PL) and reflection spectra of encapsulated monolayer molybdenum diselenide (MoSe$_2$) at various carrier density (up to $\sim 1.3 \times 10^{13} cm^{-2}$). At low doping level we observe the charge bound states of the band edge exciton A1s, its first excited state A2s, and the B exciton of the interband transition between spin-split bands (fig. 1c) B1s. With the increased carrier density ($n_e \gtrsim 2 \times 10^{11} cm^{-2}$ or $n_h \gtrsim 1 \times 10^{12} cm^{-2}$), the optical responses show distinct features of Fermi polaron in both A and B exciton series. The excited state A2s polaron characterizes much larger energy splitting between repulsive and attractive polarons due to stronger interactions. At further higher carrier density (n $\gtrsim 5 \times 10^{12} cm^{-2}$), the results deviate from Fermi polaron picture, suggesting the possibility of the complex disassociation due to the highly degenerated Fermi sea. Our work quantitatively demonstrates the regime where many-body effect predominates optical properties in monolayer TMDs and raises the need of a more sophisticated microscopic model to describe the many body behaviors of the Boson-Fermi sea complex.

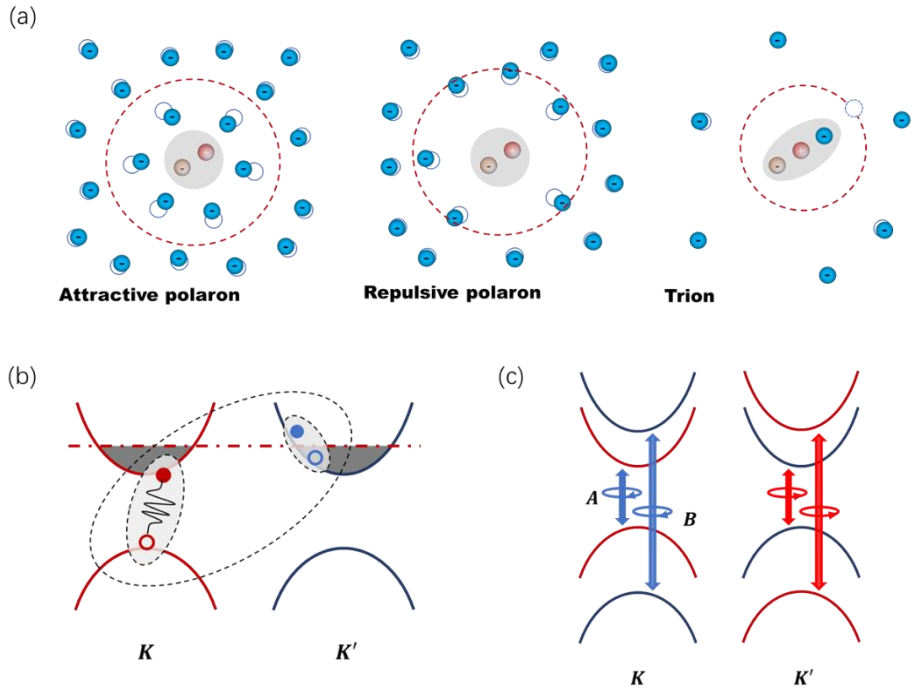

Figure 1. (a) Schematic diagram of attractive (left), repulsive (middle) exciton-polaron in the Fermi sea and trion at extremely dilute electron gas in real space. (b) Schematic diagram of exciton-polaron in k-space. (c) A diagram showing the prominent optical transitions in monolayer MoSe$_2$. A and B represent the two interband optical transitions of spin split bands (red for spin up and blue for spin down), respectively.

The h-BN encapsulated monolayer MoSe$_2$ is fabricated with dry transfer method and few-layer graphene are exploited as electric contact and electrostatic top gate.[28] The h-BN serves as both MoSe$_2$ encapsulating cover and dielectric layer. The reflection and PL spectra of samples measured at high doping level (sample 3 that $V_G$ varies from $-20V$ to $17V$) are collected in vacuum at 77K and other data are collected at 14K. PL spectrum is recorded under a 532nm continuous wave laser and reflection spectrum is measured with a tungsten lamp. The excitation laser intensity is kept low, ~100μW, to maintain the minority role of exciton in the system ($\sim 10^{10} cm^{-2}$).

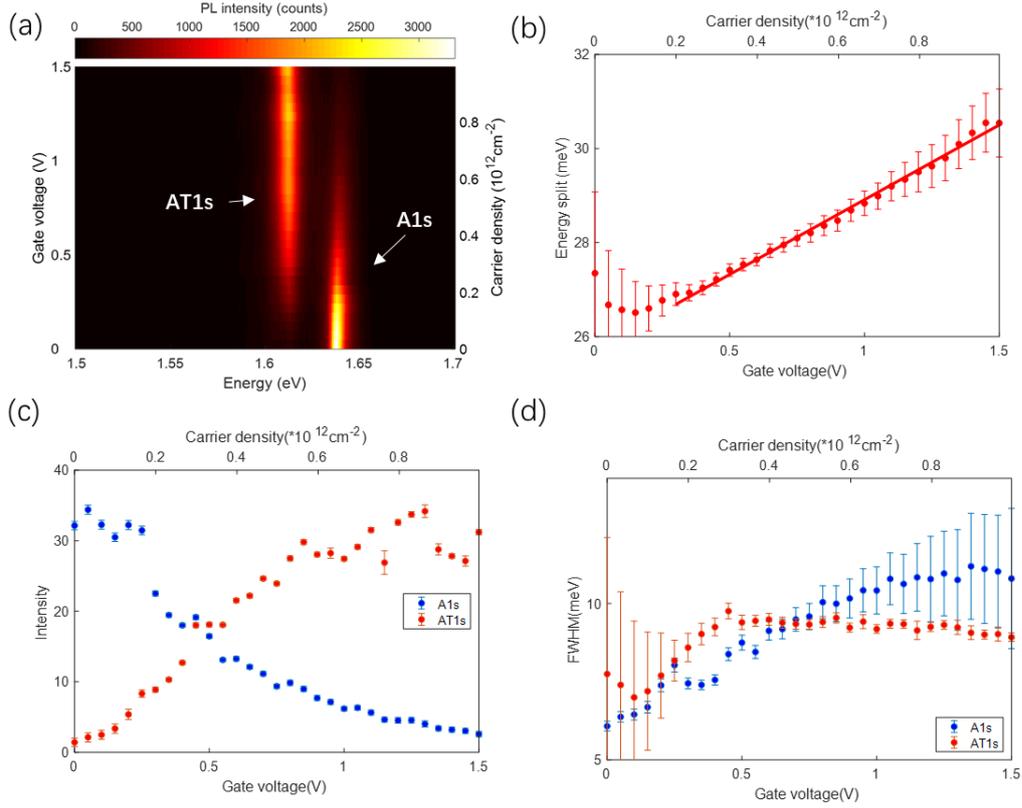

Figure 2. (a) Gate dependent PL map of sample 1 across A 1s/A trion 1s state measured at neutral and electron doping regions. (b) Energy separation between A1s and AT1s as a function of the gate voltage/carrier density. Carrier density is calculated with a parallel plate capacitor model (SI). The red solid line displays a linear fitting. (c, d) PL intensity and peak full width at half maximum (FWHM) of A1s and AT1s as a function of the gate voltage/carrier density.

**Trion to polaron picture**. The band edge exciton (labeled as A1s) and its charge bound state (AT1s), are recognized in PL map shown in fig. 2a. AT1s resides around 29meV at the lower energy side of A1s around 1.637eV near charge neutral point. The PL emission intensity transfers from A1s to AT1s upon the rise of electron density. The energy splitting (Δ) between A1s and AT1s at low electron density ($V_G < 0.3V$) shows a slight decrease trend. These are consistent with the tightly binding electron-exciton model of trion where background electrons play a very limited role in energy shift by compensating Coulomb screening and bandgap normalization.[29] However, the central peak energies of exciton A1s and trion AT1s deviate as $n_e \gtrsim 2 \times 10^{11} cm^{-2}$. Typically, the free excitons A1s tend to blueshift whereas the trions tend to redshift with the increase electron density. Apart from that, the fast emission intensity transfer from A1s to AT1s (as well as in reflection spectrum in fig. S3) and the peak linewidth change are also puzzling when the trion picture is applied. The linewidth of A1s peak broadens with elevated carrier density, while the AT1s peak width maintains within medium doping level, which agrees with the theoretical prediction of the Fermi polaron model in ref. [26] For the sake of consistency with previous literatures, the repulsive and attractive polarons evolving from excitons state at low doping region are still labelled as exciton and trion, respectively. In the polaron picture, an exciton is considered as a Boson impurity dressed by the electrons in Fermi sea via Coulomb interaction and Pauli exclusion. The repulsive and attractive polaron could be recognized as the antibonding and bonding states of excitons and the Fermi sea electron-hole pairs as illustrated in fig. 1a.[26, 30] Fig. 2b shows the energy splitting Δ is almost linearly dependent on electron density (or on $\epsilon_F$ assuming the constant density of states in 2D). The slope is deduced to be ~1, which is in good agreement with reported theoretical predictions.[26] The repulsive polaron which is an excited state of attractive polaron experiences a thermodynamically instability when the repulsive interaction increases with the increasing carrier density.[21] This instability explains the vanishing A1s emission and enhanced AT1s emission shown in fig. 2c. A broader peak at higher carrier density of the repulsive branch is also observed in fig. 2d, while the attractive branch maintains its peak width.

**Polarons associated with ground or excited excitonic states.** Fig. 3 summarizes a PL map under various electron/hole doping. Besides the prominent peaks of band edge exciton A1 and AT1s, the narrower peak at lower energy (1.786eV) is assigned to the excited state of A exciton, labelled as A2s, and the bright peak centered at 1.833eV is assigned to the spin-off B exciton. The two peaks are confirmed with the reflection (SI) and PL spectra at various electric doping level. As carrier density is lifted by external gate bias in either electron or hole side, an adjacent emission peak at ~25meV lower side to the excited state A2s emerges, which is assigned to the charge bound state of the A2s exciton, labeled as AT2s. The AT2s state rises from none at electrical neutral position and rapidly redshifts as the gate bias bilaterally increases. The charge bound state of B exciton is also observed at ~22meV lower energy side to B1, labelled as BT1s in fig. 3b. It is further confirmed by reflection spectrum in fig. 4.

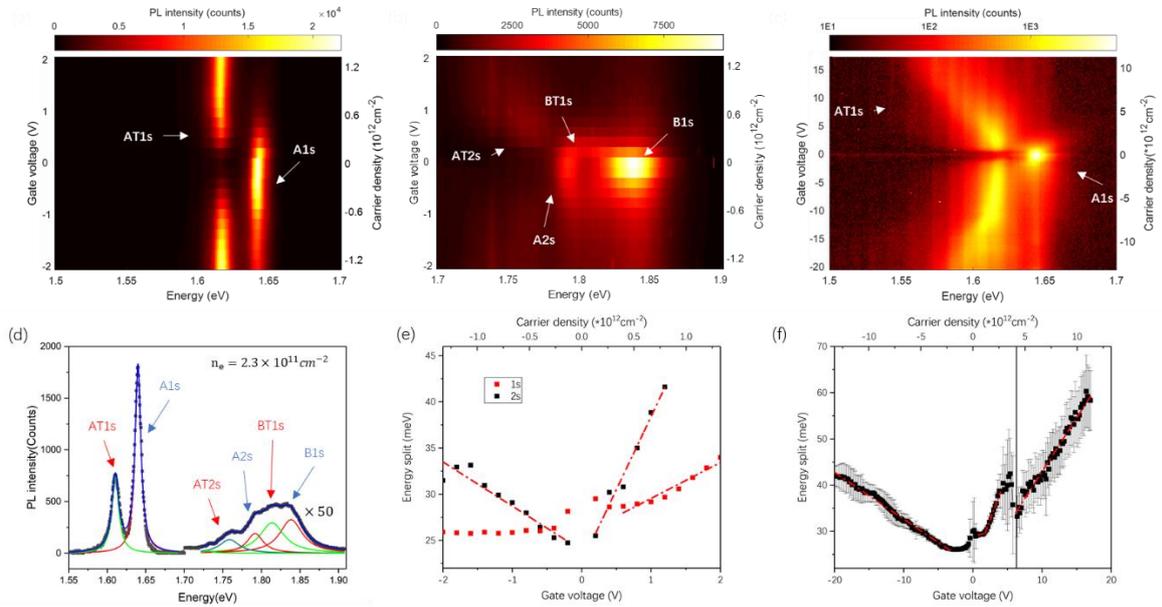

Figure 3. Gate dependent PL maps both the electron and hole doping sides across A 1s/A trion 1s (a) and B 1s, B trion 1s and A 2s, A trion 2s states (b), respectively. The negative numbers represent hole density. (c) Gate dependent PL map at 77K. For clarity, the figure is plotted in log scale. (d) Representative PL spectrum at $n_e = 2.3 \times 10^{11} cm^{-2}$ (grey dots) The spectrum could be well fitted with a superposition of a few Lorentz shaped emissions (solid lines). PL intensity near B exciton emission (1.7~1.9eV) is multiplied by 50 for legibility. (e) Energy separation (peak to peak) of A1s to AT1s and A2s to AT2s as a function of gate voltage. The red solid lines present linear fits. (f) Energy separation between A1s and AT1s as a function of gate bias deduced from (c). The vertical black line indicates electrons start to occupy the upper spin split conduction band.

Fig. 3e and f present the energy splitting $\Delta$ between A1s and AT1s and between A2s and AT2s as a function of gate bias. The linear dependence on the carrier density (Fermi energy) is observed at both $\Delta_{A1s-AT1s}$ and $\Delta_{A2s-AT2s}$ except for $\Delta_{A1s-AT1s}$ at low hole doping side ($n_h < 3 \times 10^{12} cm^{-2}$ or $\epsilon_F < 10 meV$). The $\Delta_{A1s-AT1s}$ keeps unchanged at low doping level and starts to rise linearly with carrier density above $n_h = 3 \times 10^{12} cm^{-2}$ ($\epsilon_F = 10 meV$). A linear fit of the energy splitting as a function of carrier Fermi energy (electron/hole effective mass=0.5/0.6.[31] See details in SI) yields slopes ~1.2 and ~0.4 for the $\Delta_{A1s-AT1s}$, ~4 and ~1.8 for the $\Delta_{A2s-AT2s}$ at electron- and hole-doped regimes, respectively. The Fermi energy dependence of the energy splitting $\Delta$ for 2s exciton state is 3~4 times larger than that for 1s state, implying a stronger interaction between the 2s exciton and Fermi sea. The distinct larger $\Delta_{A2s-AT2s}$ is strongly against the trion picture,[24] while it is intuitively reasonable in the polaron picture as 2s exciton characterizes a larger Bohr radius and 2s type of exciton envelope wave function (vs. 1s type for ground state A1s). Actually, the interaction $U_{QKq}$ (eq. 1) in polaron model has a non-local nature. Dmitry K. Efimkin shows this non-local interaction can be well substituted by a doping dependent local counterpart determined only by transferred momentum **q**, indicating the valid approximation of charge-dipole interactions. Here the second-order perturbation calculation is used and yields that the interaction energy $U$ of exciton 2s state is about 5.4 times of that of 1s state under a finite electric field which is consistent well with our results (SI).

The similar trend is also observed in B 1s state. The PL and reflection spectra in fig. 3b and fig 4 show that the oscillator strength transfers from B1s to BT1s with increasing carrier density. A slight blueshift of B1s emission could be recognized in the PL spectrum (fig. 3b) BT1s shows obvious redshift shown in fig. 4. The contrasting energy shift is in consistent with the Fermi polaron picture.

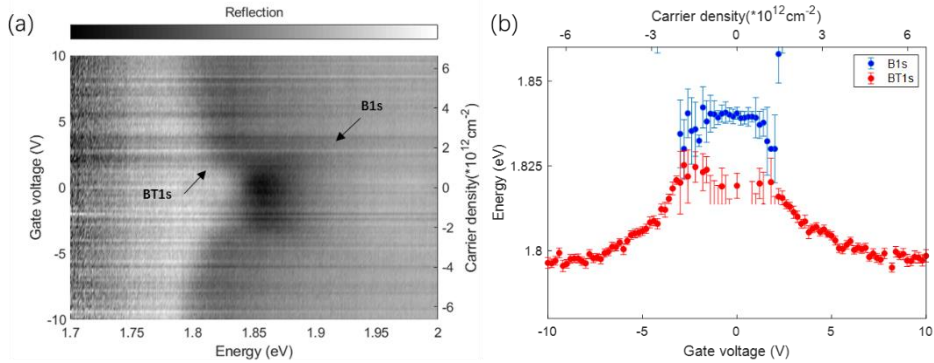

Figure 4. (a) Reflection spectrum as a function of gate bias. (b) Energy peaks of B1s and BT1s which are deduced from absorption spectrum (SI) calculated with Kramers-Kronig relation plotted as a function of the gate bias.

**Spin splitting in conduction band.** There is an abrupt energy change of A1s exciton at $\epsilon_F \approx 20 meV$ ($V_G \approx 6.3V$), which is attributed to the onset of electrons filling in the upper spin split conduction band. When the Fermi level is lifted to the upper spin split band, the fermi surface will cross four bands (spin up and down bands in two opposite valleys) as illustrated in fig. 5a, leading to an abrupt change in the density of states. An additional peak appears between A1s and AT1s as $\epsilon_F$ exceeds the threshold as shown in fig. 5b at electron doping side, while the hole doping side shows no obvious additional peak. We attribute this newly emerging peak to attractive polaron induced by the effective occupation of the upper spin split conduction band (see fig. S5 in SI), analogous to the inter- and intravalley configurations of electron-exciton in the tightly binding trion model. In the traditional trion model, intra and intervalley trions could coexist once the spin split conduction is filled and these two kinds of trions have slightly different energy owing to exchange interaction. While in Fermi polaron picture, Pauli exclusion is lifted when the exciton interacts with electrons around Fermi surface when the Fermi level is inside the upper spin split band.[32, 33] The exchange interaction could further split the attractive mode into two distinct modes depending on the electron-exciton alignment.[34] These phenomena do not come up at hole-doped regime owing to the large spin splitting of ~180meV in valence band which is not possible to fill in the lower spin split valence band with the current geometry.[31]

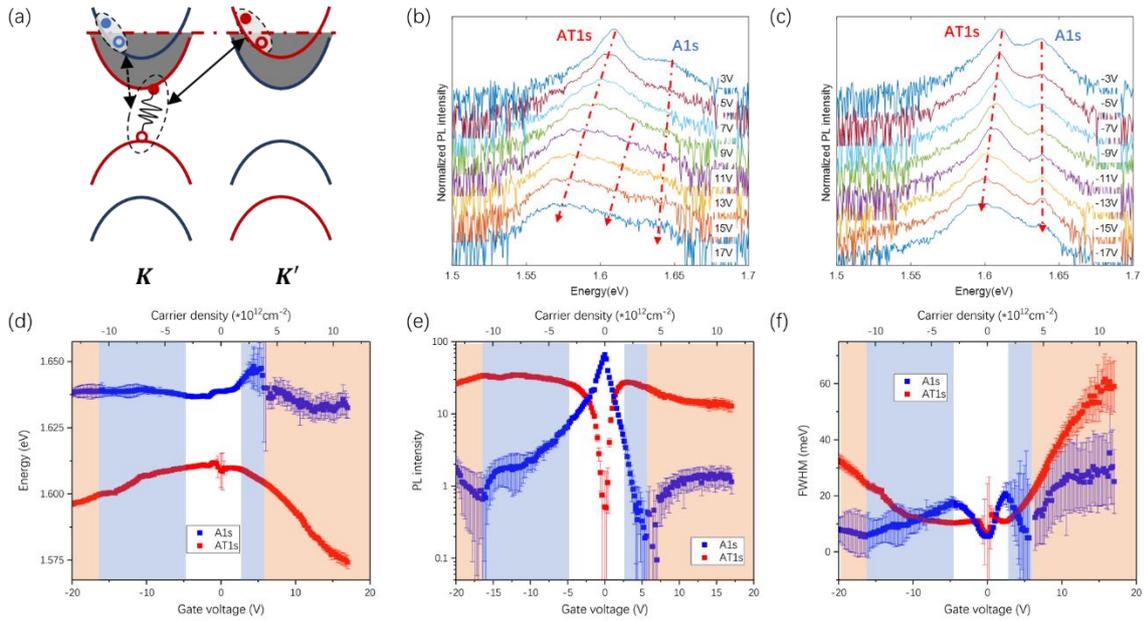

Figure 5. (a) Schematic diagram of exciton interacting with intervalley electrons in upper or lower spin split conduction bands. (b, c) Representative PL spectra at various gate biases at 77K plotted in log scale. The dashed lines act as eye guidance to indicate the evolution of the peaks. (d-f) A1s and AT1s peak energies, integrated PL intensity and FWHM deduced by fitting data in fig. 2c with Lorentz function, respectively. As mentioned in the main text, AT1s is consist of two peaks as $n_e > 3.3 \times 10^{12} cm^{-2}$. We only plot the properties of the predominant peak which is observed starting from very low doping. The white area indicates the low and medium doping region explained by polaron picture. The orange area is high doping region and the white part is transition region. Blue shaded area indicates the intermediate zone.

Fig. 5d-f summarize the PL peaks at high doping density regime. A1s and AT1s deviate in energy as $\epsilon_F$ increases. The intensity of AT1s increases with doping level at first, but starts to saturate and then decrease as $n_h > 8 \times 10^{12} cm^{-2}$ at hole doping side and $n_e > 2 \times 10^{12} cm^{-2}$ at electron doping side. Correspondingly, the FWHM of AT1s and A1s also experience non-monotonic dependence on gate bias. The exciton A1s bandwidth increases at low electron/hole density, then decreases and eventually starts to increase at high electron/hole

density regimes. While the bandwidth AT1s is insensitive to carrier density at low-medium doping regime, and monotonically increases at high electron/hole density. The AT1s width is in the same magnitude of carrier Fermi energy and the broadening may result from enhanced scatterings at a wide momentum/energy spectrum of the Fermi sea. It is noted that, as $\epsilon_F$ gets near to exciton binding energy, the assumption of exciton as atomic impurity does not hold where the carrier spacing is comparable with the exciton Bohr radius. The interactions between exciton and the Fermi sea should be re-examined microscopically.

Upon sufficiently high carrier density, the carrier Fermi energy is comparable with exciton binding energy since the enhanced screening effect could lead to a significant decrease in exciton binding energy. Therefore, the electron's kinetic energy could complete with the Coulomb attraction towards ionization of excitons (electron-hole plasma). The carrier density towards Mott criterion is expected to be $\sim 1 \times 10^{13} cm^2$ and even lower for monolayer MoSe$_2$ in dielectric environment.[35] In such a carrier density regime, Mahan exciton or Fermi-edge singularity,[36, 37] an excitonic bound state of one electron(hole) bound to the ensemble of degenerated holes(electrons) may contribute to the transitions we observed. Despite the lack of clear evidence, the increase of exciton A1s bandwidth and its PL weight at high carrier density (labeled as orange region in fig. 5) may be related to Fermi-edge singularity.

**Asymmetric behavior at electron and hole doping.** The remarkable asymmetric behavior of A exciton series at electron vs. hole doping side could not be attributed to the finite difference between effective mass of electrons (0.5m$_e$) and holes (0.6m$_e$).[31] At the elevated carrier density, the energy splitting between exciton and trion peak (fig. 3e, f and fig. 5d), the broadening of peaks (fig. 5f) and the reduction of A exciton/trion emission intensity ratio (fig. 5e) evolve with asymmetric pace at electron and hole sides (slower at hole doping side). However, B polaron redshifts at roughly symmetric pace both at electron and hole sides, in contrast to that of A trion shown in fig. 4. This finding, as well as the observed electric neutral point locating at zero-gate voltage, eliminates the possibility of unintentional n-doping, surface charge traps or formation of Fermi level pinning during device manufacturing. Here we speculate a possibility that the small spin splitting in conduction band influences A exciton series behavior, but not the B exciton which consists of electron at the upper conduction band and hole at the lower valence band, either by exchange interaction between electrons or short-wave plasmon emission with dense electron doping.[38, 39] At the hole doping side, the lower spin-split valence band barely affect the A exciton behavior due to its large energy gap with upper valence band.

In summary, we report the charge bound exciton states of the band edge exciton, its excited state and the spin-off excitons under various carrier density. The contrasting energy shift between exciton and charge bound excitons (repulsive and attractive polaron modes) and the remarkably different gate dependence of the polaron energy splitting between the ground state and the excited state excitons unambiguously support the Fermi polaron picture for excitons in monolayer TMDs. The asymmetric role of electron vs. hole in polaron and the mechanism at high carrier density remain unclear.

During the preparation of the manuscript, we are aware of two similar works.[40, 41]

**Acknowledgement:**
We thank Wang Yao for fruitful discussion and Dmitry Efimkin for communicating his unpublished result on theoretical interpretation. This work is supported by RGC General Research Fund (GRF- 17304518). Z. L. acknowledges the support from National Research Foundation Singapore programme NRF-CRP21-2018-0007, NRF-CRP22-2019-0007 and Singapore Ministry of Education via AcRF Tier 3 Programme 'Geometrical Quantum Materials' (MOE2018-T3-1-002).

**Author contribution:**
The idea was conceived by T. Y. and X. C. The experiment was carried by T. Y. and K. X. The experiment data was analyzed by T. Y. The crystal is grown by R. D. and Z. L. The device was fabricated by K. X., Q. L., T. Y. and S. Y. The theoretical analysis was performed by K. X. and C. K. The manuscript was written by T. Y., X. C and K. X. with the aid of all co-authors.

**Supplementary information**

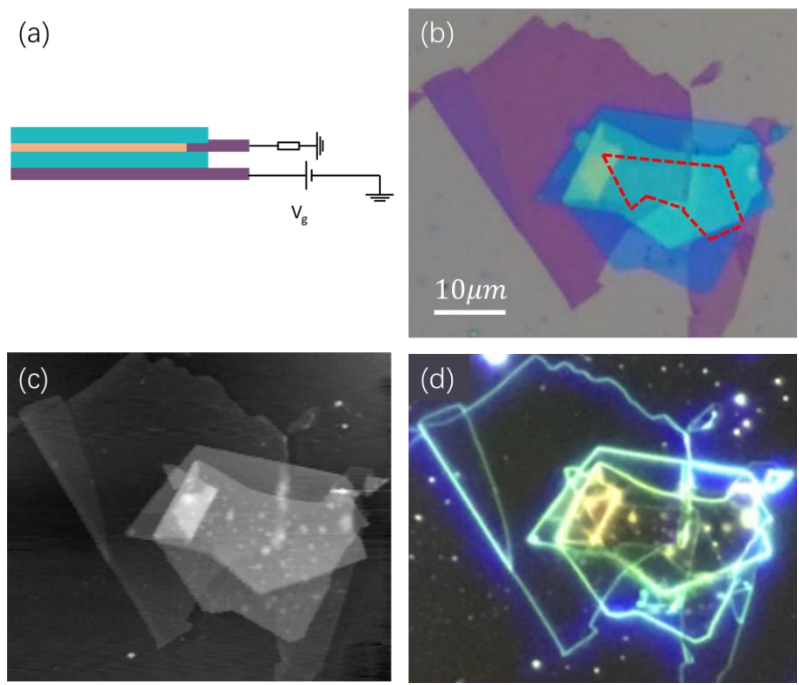

Figure S1. (a) Schematic structure of the device used in the experiment. Green rectangles represent h-BN, which cover monolayer MoSe$_2$ on both sides. The orange part is MoSe$_2$ and the purple parts are graphite. (b) Microscopic image for a typical sample on SiO$_2$/Si substrate, red dashed line represents the shape of monolayer MoSe$_2$. (c, d) Atomic force microscope and dark field optical microscope image of the same sample, respectively.

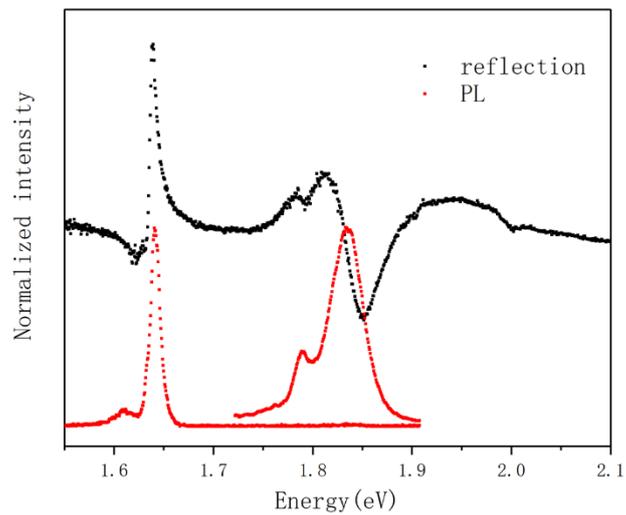

Figure S2. Photoluminescence and reflection spectra measured at $V_G = 0V$..

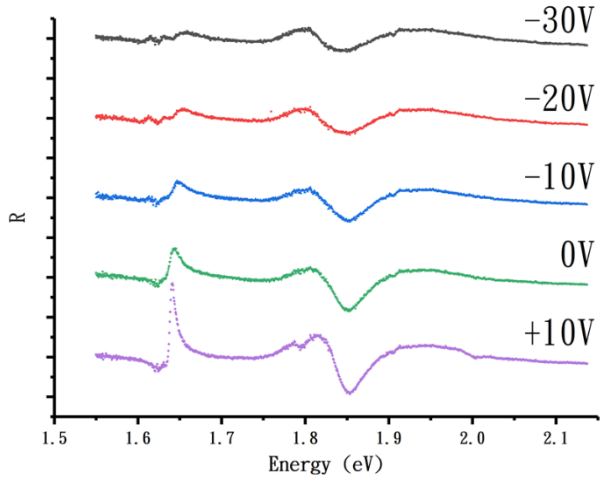

Figure S3. Reflection spectra of monolayer MoSe$_2$ collected at various gate voltages. The sample is charge neutral at $V_G \sim 10V$.

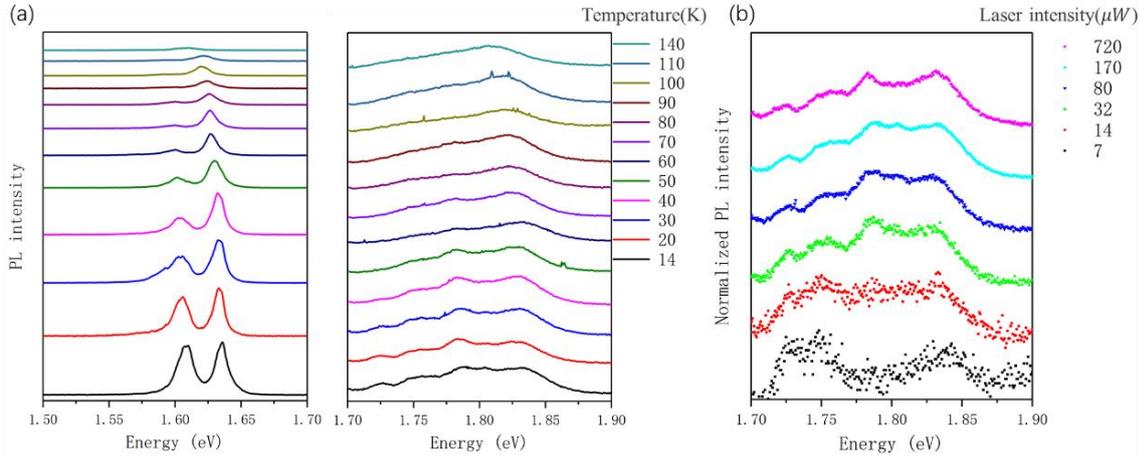

Figure S4. (a) Temperature-dependent PL spectra of monolayer MoSe$_2$ with features near A exciton and B exciton plotted in left and right figure, respectively. (b) Excitation power-dependent Pl spectra near B exciton.

Carrier density and Fermi energy calculation

The lateral dimension of graphene-BN-MoSe$_2$-graphene sample structure is typically in the unit of micrometer. The thickness of h-BN $d$ which acts as dielectric layer is $\sim 25nm$, way smaller compared to the lateral dimension. The structure is thus considered as a parallel-plate capacitor. The charge density $n$ induced by a given voltage $V$ is $\epsilon V/de$. The quantum capacitance of MoSe$_2$ $C_q = e^2 D(E)$ is neglected due to existence of optical doping.

The Fermi energy of carrier could be calculated in the simplest parabolic approximation of band structure. $E_F = \frac{n}{D(E)} = \frac{n\pi\hbar^2}{m^*}$, in which $D(E)$ is density of states in two-dimension. We take the effective mass calculated in ref. [Kormányos, A., et al. (2015). 2(2): 022001]. $m_e^1 = 0.58 m_e$, $m_e^2 = 0.5 m_e$ and $m_h^1 = -0.6 m_e$. The spin split in valence band energy is considered to be in the order of $200 meV$. The current structure can not afford for the gate voltage large enough to provide the doping hole density to fill the spin split valence band. The spin split of the conduction band is $\sim 20 meV$ equaling to the electron density doped with a gate voltage of ~6.3V.

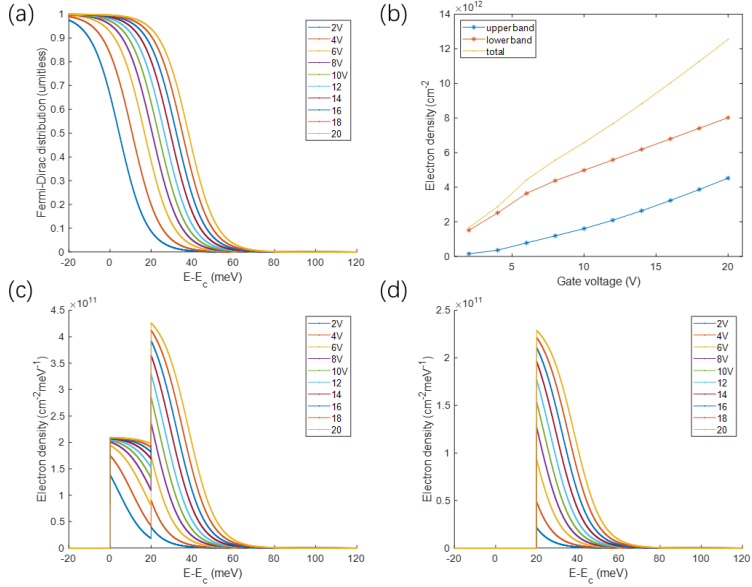

Figure S5. (a) Electron Fermi-Dirac distribution at 77K calculated at different gate voltage. (b) Electron density at the upper spin split conduction band (blue), lower band (orange) and total (yellow) as functions of gate voltage. (c) Density of electrons in the conduction band as a function of energy (E-$E_C$) plotted at varies gate voltages. (d) Density of electrons in the upper spin split conduction band as a function of energy (E-$E_C$) plotted at various gate voltages.

**Interaction strength difference between 1s and 2s state:**

the attraction force between an exciton and an electron can be considered classically: a dipole located in the electrical field induced by an electron, where a dipole moment is $\vec{\wp} = \alpha \vec{E}$, $\alpha$ is exciton polarizability, $\vec{E}$ is the electrical field induced by an electron. [Efimkin, D. K. and A. H. MacDonald (2017). Physical Review B 95(3): 035417.]

Perturbation theory is used to determine the polarizability $\alpha$ of 1s state and 2s state.

$$H_1 = -e\vec{r}\vec{E}$$

For 1s state, since the s state wave function has a central symmetry, the first order of perturbation is zero: $V_1 = \langle 0,0|H_1|0,0\rangle = 0$. The second-order term can be written as:

$$V_2^{1s} = \sum_{n,m} \frac{|\langle 0,0|H_1|n,m\rangle|^2}{\varepsilon_X^{00} - \varepsilon_X^{nm}} + \sum_{\vec{p}} \frac{|\langle 0,0|H_1|\vec{p}\rangle|^2}{\varepsilon_X^{00} - \varepsilon_X^{\vec{p}}}$$

The first term is virtual transitions from initial state to other state, the second term describes virtual ionization transitions. We get

$$V_2^{1s} = -0.1973 \frac{E^2 e^2 a_B^2}{\varepsilon_X^{00}} = -\frac{\alpha_{1s}}{2} E^2 \quad \text{with} \quad \alpha_{1s} = 0.4 \frac{e^2 a_B^2}{\varepsilon_X^{00}}$$

For estimation of 2s state, nondegenerate perturbation theory is still used since 2s and 2p states are not degenerate taking screening effect into account, assuming the energy difference between 2s and 2p state is about 50 meV. As a matter of fact, for 2s state, the excited states such as 3s state, 3p states, etc. merge with the continuum and only the second term survives.

$$V_2^{2s} = \sum_{n,m} \frac{|\langle 1,0|H_1|n,m\rangle|^2}{\varepsilon_X^{10} - \varepsilon_X^{nm}} + \sum_{\vec{p}} \frac{|\langle 1,0|H_1|\vec{p}\rangle|^2}{\varepsilon_X^{10} - \varepsilon_X^{\vec{p}}}$$

We get

$$V_2^{2s} = -\frac{\alpha_{2s}}{2} E^2 \text{ with } \alpha_{2s} = 11 \frac{e^2 a_B^2}{\varepsilon_X^{10}}$$

$$E = \frac{e}{4\pi\kappa R^2}$$

By simple regularization:

$$V_{reg}^{1s}(R) = -\frac{\alpha_{1s}}{2}\left(\frac{e}{4\pi\kappa}\right)^2 \frac{1}{(R^2 + a_{B1s}^2)^2}$$

$$U_{1s} = V_{reg}(q = 0) = \frac{\pi\alpha_{1s}}{2}\left(\frac{e}{4\pi\kappa}\right)^2 \frac{1}{a_{B1s}^2}$$

The ratio $\gamma$ of 1s state to 2s state binding energy is about 2~4 taking the screening effect into consideration. [He, Keliang, et al (2014). Physical review letters 113(2): 026803.], [Goryca, M., et al (2019). Nature communications 10(1): 1.] $t$ is introduced to account for the ratio between the Bohr radius of 2s state and 1s state, here we take $\gamma = 3$, $t = 3.9$.

$$\frac{U_{1s}}{U_{2s}} = \frac{\alpha_{2s}}{\alpha_{1s}}\frac{1}{t^2} \approx 28 \frac{\varepsilon_X^{00}}{\varepsilon_X^{10}}\frac{1}{t^2} \approx 5.4$$